\documentclass[12pt,twoside]{article}
\usepackage{fleqn,espcrc1,amssymb,epsfig,psfig}
\usepackage{graphicx}

\newcommand{\AmS}{{\protect\the\textfont2
  A\kern-.1667em\lower.5ex\hbox{M}\kern-.125emS}}

\title{Binding Energies and Scattering Observables
       in the $^4$He$_3$ Atomic System\thanks{This work was
supported by the Deutsche Forschungsgemeinschaft, Russian Foundation for
Basic Research, and National Research
Foundation of South Africa.\newline Contribution to Proceedings 
of 16th International Conference on Few-Body Problems in Physics,      
6-10 March 2000, Taipei, Taiwan. LANL e-print physics/0009035.}}

\author{A. K. Motovilov\address{JINR, 141980 Dubna, Moscow Region, Russia},
W. Sandhas\address{PI Universit\"at Bonn,
Endenicher Allee 11-13, D-53115 Bonn, Germany},
S. A. Sofianos\address{Physics Department,
UNISA,  P.\,O.\,Box 392, Pretoria 0003, South Africa}, and
E. A. Kolganova$^{\rm a}$\address{IAMS, Academia Sinica, PO Box 23-166, Taipei,
Taiwan 10764, R.\,O.\,C.}}
\begin{document}

\maketitle

\begin{abstract}
The  $^4$He$_3$  system is investigated using a hard-core 
version of the Faddeev differential equations and realistic 
$^4$He--$^4$He interactions. We calculate the binding energies 
of the $^4$He trimer but concentrate in particular on scattering 
observables.  The atom-diatom scattering lengths are calculated 
as well as the atom-diatom phase shifts for center of mass 
energies up to $2.45$\,mK.
\end{abstract}
\bigskip

\medskip

There is a great number of experimental and theoretical studies
of the $^4$He three-atomic system (see, e.\,g.,
\cite{GrebToeVil,DimerExp1,Gloeckle,Barnett,CGM,%
EsryLinGreene,Lewerenz,Nielsen,RoudnevYakovlev} and references
cited therein).  Most of the theoretical investigations
consist merely in computing the bound states, while 
scattering processes found comparatively little attention.  In
Ref.~\cite{Nakai} the characteristics of the He--He$_2$
scattering at zero energy were studied. The recombination
rate of the reaction $(1+1+1\to2+1)$ was estimated in
\cite{Fed96}. The phase shifts of the He--He$_2$ elastic
scattering and breakup amplitudes at ultra-low energies have
been calculated for the first time just recently
\cite{KMS-JPB} but only for the comparatively old HFD-B
potential by Aziz {\it et al.}~\cite{Aziz87}.

In the present paper we extend the investigations of Ref. 
\cite{KMS-JPB}. There, the formalism, which consists of a 
hard-core version of the Faddeev differential equations, has 
been described in detail.  As in \cite{KMS-JPB} we use the 
finite-difference approximation of the two-dimensional 
partial-wave Faddeev equations.  We consider only the case of 
zero total angular momentum and take 
$\hbar^2/m=12.12$\,K\,\AA$^2$. In this work, we employ grids of 
the dimension 500--800 in both the hyperradius and hyperangle, 
while the cutoff hyperradius is chosen to be up to 1000\,{\AA}.  
As compared to \cite{KMS-JPB} we use in the present work the 
refined He--He interatomic potentials LM2M2 of Ref. 
\cite{Aziz91}, and TTY of Ref. \cite{Tang95}. Our numerical 
methods have also been substantially improved, which allowed us 
to deal with considerably larger grids.  Furthermore, due to 
better computing facilities, we could take into account more 
partial waves.

Although we have performed detailed calculations of the $^4$He$_3$
binding energies, the main goal of this work was to perform
calculations for the scattering of a helium atom off a helium dimer
at ultra-low energies. Our results for the trimer binding
energies and $^4$He--atom  $^4$He--dimer scattering lengths for
the potentials employed are presented in Table~\ref{Table}. We also put
in this table the corresponding dimer binding energies together
with the $^4$He--$^4$He atomic scattering lengths. It should be noted
that the main contribution to the trimer binding energies
stems from the $l=0$ and $l=2$ partial-wave components, the
latter being about 30\,\%, and is approximately the same for all
potentials used.  The contribution from the $l=4$ partial wave
is of the order of 3-4\,\% (cf.~\cite{CGM}). We notice that that
our results for the ground-state energy $E_t^{(0)}$ of the
trimer for $l_{\rm max}=4$ are in a perfect agreement with the
corresponding values obtained in the most advanced calculations
\cite{Barnett,Lewerenz,Nielsen,RoudnevYakovlev}. The same also
holds true for the excited-state energy $E_t^{(1)}$. Our results
for $E_t^{(1)}$ are in quite a good agreement with those
of Refs. \cite{Nielsen,RoudnevYakovlev}.


\begin{table}
\caption{Dimer energies $\epsilon_d$, $^4$He$-$$^4$He diatom
scattering lengths $\ell_{\rm sc}^{(2)}$, trimer ground-state
energies $E_t^{(0)}$, trimer excited-state energies $E_t^{(1)}$,
and $^4$He atom -- $^4$He dimer scattering lengths $\ell_{\rm
sc}$ for the potentials used.}
\centering
\label{Table}
\begin{tabular}{|c|cc|cccc|}
\hline
Potential & $\epsilon_d$ (mK)  & $\ell^{(2)}_{\rm sc}$ (\AA)
          &  $l_{\rm max}$ & $E_t^{(0)}$ (K) &
             $E_t^{(1)}$ (mK) &  $\ell_{\rm sc}$ (\AA)\\
\hline \hline
        &            &         &  0 &  $-0.0942$  &  $-2.45$ & $168$  \\
\cline{5-7}
 HFD-B  & $-1.68541$ & $88.50$ &  2 &  $-0.1277$  &  $-2.71$ & $138$  \\
\cline{5-7}
        &            &         &  4 &  $-0.1325$  &  $-2.74$ & $135$  \\
\hline \hline
        &            &         &  0 &  $-0.0891$  &  $-2.02$ & $168$  \\
\cline{5-7}
 LM2M2  & $-1.30348$ & $100.23$&  2 &  $-0.1213$  &  $-2.25$ & $134$  \\
\cline{5-7}
        &            &         &  4 &  $-0.1259$  &  $-2.28$ & $131$  \\
\hline \hline
        &            &         &  0 &  $-0.0890$  &  $-2.02$ & $168$  \\
\cline{5-7}
 TTY    & $-1.30962$ & $100.01$&  2 &  $-0.1212$  &  $-2.25$ & $134$  \\
\cline{5-7}
        &            &         &  4 &  $-0.1258$  &  $-2.28$ & $131$  \\
\hline
\end{tabular}
\end{table}

There are not many results in the literature concerning the 
He--He$_2$ scattering length. Apart from our previous result 
\cite{KMS-JPB}, there is that of Ref.~\cite{Nakai} of $\ell_{\rm 
sc}=195$\,{\AA}, obtained within a zero-energy scattering 
calculation based on a separable approximation of the oldest 
Aziz {\it et al.} potential HFDHE2, and a more recent one 
obtained by Blume and Greene \cite{BlumeGreene} via a Monte 
Carlo hyperspherical calculation with the LM2M2 potential.  The 
latter result of $\ell_{\rm sc}=126$\,{\AA} is in  good 
agreement with our result of $131\pm5$\,{\AA} (see Table 
\ref{Table}).  Within the accuracy of our calculations, the 
scattering lengths provided by the LM2M2  and TTY potentials, 
like the energies of the excited state, are exactly the same.  
It should be mentioned that in this case also the two-body 
binding energies and scattering lengths are almost indentical.


\begin{figure}
\centering
\epsfig{file=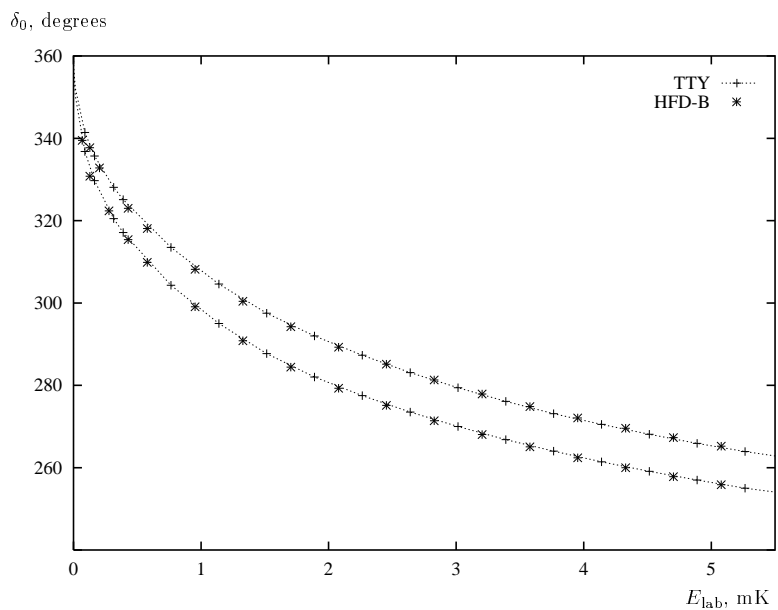,height=8cm}
\vskip -10mm
\caption{$^4$He atom -- $^4$He dimer scattering shifts for 
the HFD-B and TTY potentials. The lower curve corresponds to the 
case where $l_{\rm max}=0$ while for the upper $l_{\rm max}=2$.
}
\label{Fig-phases-black}
\vskip -2mm
\end{figure}

We have also calculated the $^4$He--atom  $^4$He--dimer 
scattering shifts for the HFD-B, LM2M2 and TTY potentials for 
center of mass energies up to $2.45$\,mK.  After transformation 
to the laboratory system the phase shifts for the these 
potentials turn out to be practically the same, especially those 
for LM2M2 and TTY.  Thus, in Fig.\,\ref{Fig-phases-black} we 
only plot the phase-shift results obtained for the HFD-B and TTY 
potentials. Note that for the phase shifts we use the 
normalization required by the Levinson theorem.  Inclusion of 
the $l=4$ partial-wave channel only adds about 0.5\,\% to the 
phase shifts obtained for $l=0$ and $l=2$. This is the reason 
why the corresponding curve for $l_{\rm max}=4$ is not depicted 
in Fig.\,\ref{Fig-phases-black}.

A detailed exposition of the material presented, including 
tables for the phase shifts, is given in an extended 
paper~\cite{He3-Bonn}.


\end{document}